\documentclass{epl}
\usepackage{graphics}
\usepackage{color}
\usepackage[colorlinks=true]{hyperref}
\usepackage{hyperref}
\title{About quantum fluctuations and holographic principle in \mth{(4+n)}-dimensional spacetime}
\shorttitle{About quantum fluctuations and ...}
\author{P. Midodashvili\thanks{E-mail: \email{pmidodashvili@yahoo.com}}}
\institute{Tskhinvali State University, 2 Besiki Str., Gori 1400,
Georgia} 
\begin{document}
\maketitle
\begin{abstract} In the article we present explicit expressions for quantum fluctuations of
spacetime in the case of $(4+n)$-dimensional spacetimes, and
consider their holographic properties and some implications for
clocks, black holes and computation. We also consider quantum
fluctuations and their holographic properties in ADD model and
estimate the typical size and mass of the clock to be used in
precise measurements of spacetime fluctuations. Numerical
estimations of phase incoherence of light from extra-galactic
sources in ADD model are also presented.
\end{abstract}
%%%%%%%%%%%%%%%%%%%%%%%%%%%%%%%%%%%%%%%%%%%%%%%%%%%%%%%%%%%%%%%%%%%
%%%%%%%%%%%%%%%%%%%%%%%%%%%%%%%%%%%%%%%%%%%%%%%%%%%%%%%%%%%%%%%%%%%
\section{Introduction}
In this article we investigate quantum fluctuations in the
spacetimes with extra spatial dimensions. In the first section,
following Ng Y.J. and van Dam H., we introduce explicit expressions
for spacetime fluctuations in $(4+n)$-dimensional spacetime, present
the implications for clocks and black holes and examine the
holographic properties for introduced fluctuations. Although our
derivation of the fluctuation expressions differs from that proposed
in \cite{Ng-vanDam}, the results agree with each other. In the
second section we investigate spacetime fluctuations in the
Arkani-Hamed-Dimopoulos-Dvali (ADD) model \cite{ADD-Model}. We show
that in case of two extra dimensions fluctuations on any distance in
the observable universe are small as compared with the size of
compact dimensions, but in the case of $11$-dimensional spacetime
fluctuations on the size of the observable universe become
comparable with the size of compact extra dimensions. We also
estimate the parameters of the clocks that can be used in precise
distance measurements and show that for $n=2$ clock's size is  much
less as compared with the size of extra dimensions. In this section
we also investigate holographic properties for the fluctuations, and
contrary to the conclusions of \cite{Scar-Cas}, explicitly show that
holography is not destroyed in ADD model. And finally we present
numerical estimations of the phase incoherence of light from
extra-galactic sources in ADD model.

Before proceeding to next section let us recall some known facts for
the black holes in the $(4+n)$-dimensional spacetime. The
Schwarzschild solution in $(4+n)$ dimensions has the form
\cite{Myers-Perry} (we set $c=\hbar=1$)
\begin{equation}\label{SchwarzschildSolution(4+n)}
\begin{array}{l}
ds^2  =  - \phi \left( r \right)dt^2 + \phi \left( r \right)^{ - 1}
dr^2  + r^2 d\Omega _{2 + n}^2 ~, ~~~~\phi \left( r \right) = 1 -
\left( {{{r_{S\left( {4 + n} \right)} } \mathord{\left/ {\vphantom
{{r_{s\left( {4 + n} \right)} } r}}
\right.\kern-\nulldelimiterspace} r}} \right)^{1 + n} , \\ r_{S(4 +
n)}  = \left[ {B_{\left( {4 + n} \right)} G_{\left( {4 + n} \right)}
m} \right]^{1/(1 + n)}~ ,~~~B_{\left( {4 + n} \right)} =
\frac{{16\pi }}{{\left( {2 + n} \right)A_{\left( {2 + n} \right)}
}},~~~~A_{\left( {2 + n} \right)} = \frac{{2\pi ^{{{\left( {n + 3}
\right)} \mathord{\left/ {\vphantom {{\left( {n + 3} \right)} 2}}
\right. \kern-\nulldelimiterspace} 2}} }}{{\Gamma \left[ {{{\left(
{n + 3} \right)} \mathord{\left/ {\vphantom {{\left( {n + 3}
\right)} 2}} \right. \kern-\nulldelimiterspace} 2}}
\right]}},\\\end{array}\end{equation} where $r_{S(4+n)}$ and
$G_{(4+n)}$ are Schwarzschild radius, associated with the mass $m$,
and gravitational constant in $(4+n)$-dimensional spacetime
respectively. $A_{\left( {2 + n} \right)}$ denotes the area of the
unit $(2+n)$-sphere. The $(4+n)$-dimensional Planck length, time and
mass are defined as follows
\begin{equation}\label{PlanckScale}
l_{Pl\left( {4 + n} \right)}  = t_{Pl(4 + n)}=m_{Pl\left( {4 + n}
\right)}^{ - 1}= G_{\left( {4 + n} \right)}^{1/(2+n)}
~.\end{equation}

%%%%%%%%%%%%%%%%%%%%%%%%%%%%%%%%%%%%%%%%%%%%%%%%%%%%%%%%%%%%%%%%%%%
%%%%%%%%%%%%%%%%%%%%%%%%%%%%%%%%%%%%%%%%%%%%%%%%%%%%%%%%%%%%%%%%%%%
\section{Quantum fluctuations of spacetime , clocks,
black holes and limits on computation in $(4+n)$-dimensional
spacetime} In this section, following the arguments proposed by Ng
and van Dam \cite{Ng-vanDam}, we analyze the thought experiment of
distance and time measurement in the case of $(4+n)$ dimensions,
write down explicit results for fluctuations of spacetime and
consider implications for clocks, black holes and computation.

It is well known in general relativity that coordinates do not have
any intrinsic meaning independent of observations; any coordinate
system is defined only by explicitly carrying out spacetime distance
measurements. In order to measure the distance between two points A
and B, one puts a clock at A and a mirror at B. By sending a light
signal from the clock to the mirror in timing experiment, one can
determine the distance $l$. Let $m$ be the mass of the clock. If the
clock has the initial linear position spread $\delta l_{in}$ when
the light signal leaves it, then its final position spread grows to
\begin{equation}\label{}
\delta l  = \delta l_{in} + l(m\delta l_{in})^{-1}~,
\end{equation}when the light signal returns to the clock.
The minimization of the final spread with respect to $\delta l_{in}$
gives
\begin{equation}\label{MinimalQuantumMechanical} \delta
l  \sim  2(l/m)^{1/2},\end{equation} so quantum mechanics alone
suggests a massive clock to reduce the uncertainty in distance.

Now suppose in the measuring of distance one uses the light-clock
consisting of a spherical cavity of diameter $a$, surrounded by a
mirror wall of mass $m$, between which bounces a beam of light. The
inevitable uncertainty in distance measurement caused by the clock
is
\begin{equation}\label{UncertaintyFromSizeOfClock}
\delta l \sim a=\beta r_{S(4+n)}.
\end{equation} Obviously, in order that the clock is not a black hole, one must
suppose that the dimensionless quantity $\beta > 1$. The uncertainty
(\ref{UncertaintyFromSizeOfClock}) according to
(\ref{SchwarzschildSolution(4+n)}) increases with mass of clock, so
the general relativity suggests a lightweight clock to do the
measurement.

The total uncertainty in distance measurement can be written as the
sum of (\ref{MinimalQuantumMechanical}) and
(\ref{UncertaintyFromSizeOfClock})
\begin{equation}\label{UnsertaintyTotal}
\delta l_{tot} \sim  2(l/m)^{1/2} +\beta r_{S(4+n)}.
\end{equation} The minimization of (\ref{UnsertaintyTotal}) with
respect to clock's mass gives the minimum uncertainty in distance
\begin{equation}\label{}
\delta l_{min} \sim C_{(4+n)}\beta ^{(1 + n)/(3 + n)} \left(
{ll_{Pl(4 + n)}^{2 + n} } \right)^{1/(3 + n)},~~C_{(4 + n)}  =
\left( {\frac{{(3 + n)^{3 + n} }}{{(1 + n)^{1 + n} }}B_{(4 + n)} }
\right)^{1/(3 + n)}.\end{equation} Taking the size of the clock as
small as possible (i.e. $\beta \sim 1$) and neglecting numerical
factors of order $1$, for the uncertainty in distance measurement
one gets
\begin{equation}\label{DistanceUncertainty}
\delta l \sim  \left( {ll_{Pl(4 + n)}^{2 + n} } \right)^{1/(3 + n)}.
\end{equation}
From eq.(\ref{DistanceUncertainty}) one can also estimate the
minimum measurable space distance by equating $\delta l=l$, and the
minimum measurable length is  $l_{\min } \sim l_{Pl\left( {4 + n}
\right)}$.

Considering similar gedanken experiment of measuring a time interval
$t$ and neglecting numerical factors, one easily gets an analogous
expression for the corresponding uncertainty $\delta t$ in time:
\begin{equation}\label{TimeUncertainty}
\delta t \sim  \left( {tt_{Pl\left( {4 + n} \right)}^{2 + n} }
\right)^{{1 \mathord{\left/ {\vphantom {1 {\left( {3 + n} \right)}}}
\right. \kern-\nulldelimiterspace} {\left( {3 + n} \right)}}}~.
\end{equation}

The uncertainties in distance and time can be translated into a
metric uncertainties over a distance $l$ and a time interval $t$
caused by the quantum fluctuations in the fabric of spacetime
\begin{equation}\label{MetricFluctuations}
\delta g_{\mu \nu }  \sim (l_{Pl(4 + n)} /l)^{(2 + n)/(3 + n)}
,~~(t_{Pl(4 + n)} /t)^{(2 + n)/(3 + n)}~.
\end{equation}

Using these results one can find upper bound on the lifetime of a
simple clock \cite{Ng-vanDam}. Indeed, if the time resolution of the
clock (i.e., the smallest time interval it is capable to measure) is
$\tau$, then its lifetime can be defined as the maximum measurable
time interval $T$ during which its time resolution exceeds the
corresponding time fluctuation (\ref{TimeUncertainty}). Thus
\begin{equation}\label{}
\tau  \ge \left( {Tt_{Pl\left( {4 + n} \right)}^{2 + n} }
\right)^{{1 \mathord{\left/ {\vphantom {1 {\left( {3 + n} \right)}}}
\right. \kern-\nulldelimiterspace} {\left( {3 + n}
\right)}}},\end{equation} from which for the lifetime of the clock
one gets
\begin{equation}\label{ClockLifetime}
T \le \tau \left( {{\tau  \mathord{\left/ {\vphantom {\tau
{t_{Pl\left( {4 + n} \right)} }}} \right. \kern-\nulldelimiterspace}
{t_{Pl\left( {4 + n} \right)} }}} \right)^{2 + n}.\end{equation} If
one considers a black hole as an ultimate clock, then its time
resolution is given by the light travel time across the black hole's
horizon, i.e. $\tau _{BH}\sim r_{S(4+n)}$, and from
(\ref{ClockLifetime}) the lifetime of the black hole is given by
\begin{equation}\label{}
T_{BH} \sim m_{Pl(4 + n)}^{ - 1} (m/m_{Pl(4 + n)} )^{(3 + n)/(1 +
n)},\end{equation} which is in accordance with the result previously
obtained in \cite{Arg-Dim-March}.

The relation eq.(\ref{ClockLifetime}) also can be used to put a
limit on the memory space of any simple information processor.
Indeed, the maximum number $I$ of steps of information processing
can be estimated by $I=T/\tau$, where $T$ is the lifetime of
processor with processing frequency $\nu=\tau^{-1}$, and so from
eq.(\ref{ClockLifetime}) one immediately gets
\begin{equation}\label{}
I\nu ^{2 + n}  \le \left( { t_{Pl(4 + n)}^{2 + n} } \right)^{ -
1},\end{equation} the bound which is universal in that it  is
independent of the mass, size and details of the simple computer.

And finally in this section let us investigate the holographic
properties of spasetime fluctuations in
$(4+n)$-dimensional spacetime. %According to the holographic
%principle, the number of degrees of freedom that any space region
%can contain is bounded by the surface area of the hypersurface
%enclosing the given region in Planck units, i.e.,
%$(l^{2+n}/l_{Pl(4+n)}^{2+n})$, instead of by the volume of the
%region.
Consider  a spatial region of volume $V_{(4+n)}$
('hypercube' measuring $l \times l \times l \times \cdot \cdot \cdot
\times l = l^{3 + n}$). The number of degrees of freedom
$N_{V_{(4+n)}}$ contained in the region is bound by the maximum
number of the small hypercubes that can be put inside the region.
But each side of the small hypercubes cannot be smaller than the
accuracy $\delta l$ with which can be measured each side $l$ of
taken spatial region, i.e. the side of small hypercube obeys
eq.(\ref{DistanceUncertainty}). Thus one has
\begin{equation}\label{}
N_{V_{(4+n)}}\sim l^{3+n} /\left( {\delta l } \right)^{3+n} \le
\left( {l/l_{Pl\left( {4 + n} \right)}} \right)^{2 +
n},\end{equation} and so the uncertainties in distance caused by the
quantum fluctuations satisfy the holographic counting of degrees of
freedom.

%%%%%%%%%%%%%%%%%%%%%%%%%%%%%%%%%%%%%%%%%%%%%%%%%%%%%%%%%%%%%%%%%%%%
%%%%%%%%%%%%%%%%%%%%%%%%%%%%%%%%%%%%%%%%%%%%%%%%%%%%%%%%%%%%%%%%%%%%
\section{Quantum fluctuations of spacetime and their implications
in the ADD model} In this section, by using the results introduced
above, we consider quantum fluctuations of spacetime and their
implications in the Arkani-Hamed-Dimopoulos-Dvali (ADD) model
\cite{ADD-Model} with $n$ extra spacelike compact dimensions of size
$L$ and low fundamental scale $m_{Pl((4+n))} \sim $ Tev. Within the
framework of this model the links between $4$- and
$(4+n)$-dimensional physical values can be written as follows
\begin{equation}\label{LinksBetween4and4+n}
G_{\left( 4 \right)}  = \frac{{G_{\left( {4 + n} \right)} }}{{L^n
}}~, ~~~~r_{S\left( 4 \right)} = \frac{{B_{\left( 4 \right)}
}}{{B_{\left( {4 + n} \right)} }}\frac{{r_{S\left( {4 + n}
\right)}^{1 + n} }}{{L^n }},~~~\frac{{m_{Pl\left( 4 \right)}^2
}}{{L^n }} = m_{Pl\left( {4 + n} \right)}^{2 + n}~,~l_{Pl\left( 4
\right)}^2 = \frac{{l_{Pl\left( {4 + n} \right)}^{2 + n} }}{{L^n }}.
\end{equation}

It is reasonable to assume that $L \gg l_{Pl\left( {4 + n}
\right)}$, otherwise the extra dimensions would not have a classical
spacetime structure. In this case from
eq.(\ref{LinksBetween4and4+n}) one gets $l_{Pl\left( {4 + n}
\right)} \gg l_{Pl\left( 4 \right)}$, and thus $L \gg l_{Pl\left( {4
+ n} \right)}  \gg l_{Pl\left( 4 \right)}$.

Neglecting numerical factors of order $1$ and using
(\ref{LinksBetween4and4+n}), from (\ref{DistanceUncertainty}) one
can estimate the distances at which fluctuations become comparable
with the size $L$ of compact extra dimensions, i.e. $\delta l \sim
L$,
\begin{eqnarray}\label{MaximumDistances}
l \sim L(L/l_{Pl(4 + n)} )^{2 + n}~~  \Rightarrow~~ l \sim
L(L/l_{Pl(4)} )^2.\end{eqnarray} Putting $m_{Pl(4+n)}=1~\rm{TeV}$
and using $l_{Pl(4)} \sim 10^{-35}~\rm{m}$ for the size  of extra
dimensions one gets
\begin{equation}\label{SizeOfExtraDimensions}
L \sim 10^{30/n - 19}~\rm{m},
\end{equation} and then, from (\ref{MaximumDistances}) easily can be found
\begin{equation}\label{}
l \sim 10^{90/n + 13}~\rm{m}~.\end{equation} For $n=2$ one gets $l
\sim 10^{58}~\rm{m}$, which is much greater then the size of the
observable universe $l_{\rm{universe}} \sim 10^{10}$ light-years
$\sim 10^{26}~\rm{m}$, thus in $6$-dimensional spacetime
fluctuations on any distances in the observable universe is much
less as compared  with the size of compact extra dimensions. We
notice that for $n=7$ one gets $l \sim l_{\rm{Universe}}$, i.e. in
case of $11$-dimensional spacetime the fluctuations on the distances
of the order of the size of observable universe become comparable
with the size of compact dimensions.

One can readily estimate the size of clock to be used in the
measurement of fluctuations $~a/L \sim r_{S(4+n)}/L \sim
(ll^{2}_{Pl(4)}/L^{3})^{1/(3+n)}$, from which, using
(\ref{SizeOfExtraDimensions}), even for the distances equal to the
size of the observable universe in cases $n=1$ and $n=2$, one has
$~a/L \sim 10^{-19}$ and $~a/L \sim 10^{-6}$ respectively. So in
this cases $a \sim r_{S(4+n)} \ll L$. One can also estimate the mass
of the clock $m \sim m_{Pl(4+n)}(l/l_{Pl(4+n)})^{(1+n)/(3+n)}$. In
case $l \gg l_{Pl(4+n)}$ the clock's mass satisfies the inequality
$m \gg m_{Pl(4+n)}$, i.e. the clock is well described in the
framework of unquantized gravity.

In order to discuss holographic properties of fluctuations in
distance in ADD model, one considers space region in the form of
hypercube of characteristic size $l$. If $l<L$, then
 volume of the region $V_{(4+n)}\sim
l^{3+n}$. Using (\ref{DistanceUncertainty}) and neglecting numerical
factors, for the number of degrees of freedom $N_{V_{(4+n)}}$
contained in the region one gets
\begin{equation}\label{SmallHyperCube}
N_{V_{(4+n)}}\sim  l^{3+n} /\left( {\delta l } \right)^{3+n} \le
\left( {l/l_{Pl\left( {4 + n} \right)}} \right)^{2 + n},
\end{equation}i.e. the uncertainties in distance satisfy
the holographic principle in purely $(4+n)$-dimensional form. If the
characteristic size of the hypercube $l>L$, then  its volume
$V_{(4+n)}$ (the region measures $l \times l \times l$ on the brane)
is equal to $V_{(4+n)}\sim V_{(4)}V_{EtraSpase}\sim l^3 L^n$ (in the
definition of extra space volume $V_{Extra Space}\sim L^n$  the
numerical factor of order $1$, depending on the exact form of
compact extra dimensions, is ignored). Thus, neglecting numerical
factors, from eq.(\ref{DistanceUncertainty}) for the number of bits
in this region one gets
\begin{equation}\label{BigHyperCube}
N_{V_{\left( {4 + n} \right)} }  \le V_{\left( {4 + n} \right)}
/\delta l_{min }^{3 + n}\sim l^3 L^n /(ll_{Pl(4 + n)}^{2 + n} )\sim
l^3 L^n /(L^{n}ll_{Pl\left( {4 } \right)}^{2 }) \sim
(l^2/l_{Pl\left( 4 \right)}^2),\end{equation} i.e. the holographic
principle is satisfied and effectively it has $4$-dimensional form.
Our conclusions about the holography in ADD model are different from
those obtained in \cite{Scar-Cas}, where a different bound from
gravity was used. \footnote{according to \cite{Scar-Cas} holography
is destroyed when extra dimensions are admitted; note that in
\cite{Ng-vanDam} (last reference in it) Ng Y.J. also noticed the
conclusions of \cite{Scar-Cas} as different from his own results.}
The origin of the difference we have considered in more details in
our previous letter \cite{Paul}.

In conclusion, let us numerically estimate the effects of spacetime
fluctuation in ADD model with n=2. In the first instance, in the
case $n=2$ and $L\sim 100~ \mu m $ on the size of the whole
observable universe ($ \sim 10^{10}$ light-years) the fluctuation in
the ADD model is $\delta l_{(6)} \sim 10^{-11} $ m, whereas in the
purely $4$-dimensional case the fluctuation is $l_{(4)} \sim
10^{-15}$ m \cite{Ng-vanDam}. Furthermore, according to the
classification proposed in \cite{Ng-vanDam},  the ADD spacetime foam
corresponds to the model with $\alpha = (2+n)/(3+n)$. Thus,
following \cite{Ng-vanDam}, one can also estimate the phase
incoherence of light from extra-galactic source in this model
\begin{equation}\label{PhaseIncoherence}
\Delta \phi_{(4+n)}  \sim 2\pi \left( {{{l_{Pl(4 + n)} }
\mathord{\left/ {\vphantom {{l_{Pl(4 + n)} } \lambda }} \right.
\kern-\nulldelimiterspace} \lambda }} \right)^\alpha  \left( {{l
\mathord{\left/ {\vphantom {l \lambda }} \right.
\kern-\nulldelimiterspace} \lambda }} \right)^{1 - \alpha
},\end{equation} where $\lambda$ is the wavelength of received
radiation from a celestial source located at a distance $l$ away.
Using (\ref{SizeOfExtraDimensions}) it is straightforward to write
\begin{equation}\label{}
\Delta \phi _{(4 + n)}  \sim 10^{16 - 18/(3 + n)} \left(
{{{l_{Pl(4)} } \mathord{\left/ {\vphantom {{l_{Pl(4)} } l}} \right.
\kern-\nulldelimiterspace} l}} \right)^{n/(9 + 3n)} \Delta \phi
_{(4)},\end{equation} where $\Delta \phi _{(4)}$ is the phase
incoherence in purely $4$-dimensional spacetime \cite{Ng-vanDam}.
For the active galaxy PKS1413+135, for which $\lambda \approx 1.6
\times 10^{-6}$ m and $l \approx 1.216$ Gpc $\approx 3 \times
10^{25}$ m, in case $n=2$ one gets $\Delta \phi _{(6)} \sim
10^{4}\Delta \phi _{(4)}$, i.e. the phase incoherence of light is
greater by $4$ orders in ADD model as compared with the purely
$4$-dimensional case. So it is worthwhile to estimate other possible
effects of spacetime fluctuations in ADD model in the view of
testing in experiments.

\acknowledgments The author is indebted to Ng Y.J. for valuable
remark and to Gogberashvili M. and Maziashvili M. for useful
conversations.

\end{document}